\documentclass[10pt, conference, compsocconf]{IEEEtran}
\ifCLASSINFOpdf
\else
\fi

\usepackage{multirow}

\usepackage[bookmarks=false]{hyperref}
\usepackage[american]{babel}
\addto\extrasamerican{

}
\usepackage{xargs}
\usepackage[pdftex,dvipsnames]{xcolor}
\newcommandx{\marco}[2][1=]{\todo[linecolor=red,backgroundcolor=red!25,bordercolor=red,#1]{\textbf{M} #2}}
\newcommandx{\danilo}[2][1=]{\todo[linecolor=blue,backgroundcolor=blue!25,bordercolor=blue,#1]{\textbf{D} #2}}
\newcommandx{\fabricio}[2][1=]{\todo[linecolor=OliveGreen,backgroundcolor=OliveGreen!25,bordercolor=OliveGreen,#1]{\textbf{F} #2}}
\newcommandx{\ana}[2][1=]{\todo[linecolor=Plum,backgroundcolor=Plum!25,bordercolor=Plum,#1]{\textbf{A} #2}}
\newcommandx{\ju}[2][1=]{\todo[linecolor=red,backgroundcolor=red!25,bordercolor=red,#1]{\textbf{JU} #2}}
\usepackage{subcaption}

\usepackage{graphicx}
\usepackage{setspace}
\usepackage[square,numbers]{natbib}

\begin{document}
\bstctlcite{IEEEexample:BSTcontrol}
\setlength\tabcolsep{5pt}
%
\title{Machine Learning for Performance Prediction of Spark Cloud Applications}


\author{\IEEEauthorblockN{Alexandre Maros, Fabricio Murai, \\ Ana Paula Couto da Silva, Jussara M. Almeida}
\IEEEauthorblockA{Department of Computer Science\\
 Universidade Federal de Minas Gerais, Brazil\\
\{alexandremaros,murai,\\ana.coutosilva,jussara\}@dcc.ufmg.br}
\and
\IEEEauthorblockN{Marco Lattuada, Eugenio Gianniti,\\ Marjan Hosseini, Danilo Ardagna}
\IEEEauthorblockA{Dipartimento di Elettronica, Informazione e Bioingegneria\\
Politecnico di Milano, Italy\\
\{marco.lattuada,eugenio.gianniti,\\marjan.hosseini,danilo.ardagna\}@polimi.it}
}


%


\maketitle

\begin{abstract}
Big data applications and analytics are employed in many sectors for a variety of goals: improving customers satisfaction, predicting market behavior or improving processes in public health. These applications consist of complex software stacks that are often run on cloud systems. Predicting execution times is important for estimating the cost of cloud services and for effectively managing the underlying resources at runtime. Machine Learning (ML), providing black box solutions to model the relationship between application performance and system configuration without requiring in-detail knowledge of the system, has become a popular way of predicting the performance of big data applications. We investigate the cost-benefits of using supervised ML models for predicting the performance of applications on Spark, one of today's most widely used frameworks for big data analysis. We compare our approach with \textit{Ernest} (an ML-based technique proposed in the literature by the Spark inventors) on a range of scenarios, application workloads, and cloud system configurations. Our experiments show that Ernest can accurately estimate the performance of very regular applications, but it fails when applications exhibit more irregular patterns and/or when extrapolating on bigger data set sizes. Results show that our models match or exceed Ernest's performance, sometimes enabling us to reduce the prediction error from 126-187\% to only 5-19\%.
\end{abstract}


%
\IEEEpeerreviewmaketitle

\section{Introduction}

Big data applications have become widespread in various domains, such as 
natural language processing~\cite{Hirschberg2015}, public health~\cite{Fang:2016}, and social media analytics~\cite{GHANI2018}. These applications are characterized by very heterogeneous and irregular data accesses and computation patterns, often built on the top of massively parallel algorithms. At the same time, cloud computing platforms, such as Amazon EC2, Google Cloud, and Microsoft Azure\footnote{http://aws.amazon.com, http://cloud.google.com, and\\ http://azure.microsoft.com, respectively.}, have also grown substantially in  popularity in recent years. These platforms offer virtualized environments, which allow users to dynamically adjust the allocated resources to match the application's current needs.
Therefore, they offer a suitable execution environment for the often highly distributed and variable processing requirements of big data applications.  

 For all these reasons, users and enterprises started adopting cloud services to run their applications  as a more cost-effective alternative to the traditional local server architecture~\cite{Low2011}. This type of infrastructure often relies on distributed programming platforms, such as Apache Spark. Spark is a fast and general engine for large-scale data processing whose adoption has steadily increased and which probably will be the reference big data engine for the next five to ten years\footnote{http://fortune.com/2015/09/25/apache-spark-survey}. Spark facilitates the implementation of a variety of applications, from relational data processing (e.g., SQL queries)  to machine learning algorithms \footnote{https://databricks.com/blog/2016/09/27/spark-survey-2016-released.html}.
 
 Cloud computing services often have an extensive list of possible configurations that may be allocated for an application.
 The user should choose the type of instances (processing nodes), the total number of cores, among others. These choices may drastically affect the application's execution time and thus should be carefully planned. The performance prediction of a given application at a target configuration becomes then a key task to support the proper planning and management of the available resources.
 
 There is a great body of work on performance prediction, most relying on traditional techniques such as  analytical models ~\cite{NelsonT88,lundstrom2004,tripathi2000,Ardagna:2018} and simulation~\cite{JMT,Wang2015}. Some studies have focused specifically on modeling the performance of parallel applications~\cite{DBLP:conf/ica3pp/ArdagnaBGAPR16}, more recently addressing also specific characteristics of 
 cloud environments~\cite{Ardagna:2018}.  Yet, these techniques   require detailed knowledge of the system, which is not always available, and  often rely on simplifying assumptions at the cost of losing accuracy. Thus, they are either not capable of capturing the intricacies and complexities of cloud-based big data applications or are too complex for practical use (i.e., cannot support real-time prediction).  
 
 In contrast, some recent studies have exploited supervised machine learning (ML) models for performance prediction of large systems~\cite{venkataraman2016ernest,MUSTAFA20183767, Hemingway, CherryPick}. These techniques
 are often referred to as {\it black box} solutions because they try to learn from previously collected data and make predictions without the knowledge of the system internals. Supervised ML models require a training phase in which they use experimental data coming from different configurations to learn a prediction function. The experiments to collect these training data may be time-consuming. However, model training takes place offline, and  once trained, the prediction of the learned ML model is fast and usually very accurate.

We here aim at investigating the cost-benefits of using supervised ML models in the performance prediction of Spark applications. Specifically,
given a set of features extracted from a target Spark application (e.g.,  size of the input data) and from a platform configuration (e.g.,  number of cores), we want to predict with reasonable accuracy the application execution time when running on the target platform. Our goal is to learn a prediction model based on data (i.e., sets of features and execution times) from prior executions of the same application under various configurations.


The aforementioned problem can be framed as a regression task. The ML literature includes several algorithms for solving regression  problems \cite{zaki2014data}. To our knowledge, the state-of-the-art in using ML algorithms to predict the performance of Spark applications is the work by Venkataraman {\it et al.}~\cite{venkataraman2016ernest}. The proposed solution, referred to as {\it Ernest} model, exploits only 
four  features which are functions of the dataset size and of the number of cores and relies on non-negative least squares (NNLS) regression.   

In other words, we address the following question: {\it To which extent can we offer more accurate predictions by exploiting other regression algorithms and/or other features?} To that end, we consider four classic ML techniques, namely linear regression (LR), neural network (NN), decision tree (DT), and random forest (RF) \cite{zaki2014data}, and build two different approaches. 
Our first approach, referred to as {\it black box}, relies only on features that capture knowledge available prior to the application execution, similarly to the {\it Ernest} model (but, as mentioned, using different ML methods). 
Our second approach, referred to as {\it gray box}, includes a richer set of features capturing more details of the application execution (e.g.,  number of tasks running within a single stage, their average execution time, etc.).
We evaluate both approaches, comparing them against the {\it Ernest}  model, on different scenarios, covering different application workloads and  platform configurations, and also investigating the impact of the size of the training set on prediction accuracy.

Our experimental results show that the {\it Ernest} model is able to accurately estimate the performance of very regular applications (errors are in the range 1.5-10.5\%), but its accuracy falls short when the application  presents more irregular behavior or  dataset extrapolation capabilities are required (error up to 187\%). In contrast, our  approach  is able to address these scenarios by providing highly accurate predictions (the largest error is less than 20\%). Moreover, our experiments reveal that there is no clear winner among the four ML techniques tested, and different techniques have to be explored to select the best one for the scenario. 


\section{Related Work}\label{sec:related}

The performance analysis and prediction of big data applications running on the cloud can be tackled from different perspectives. The most traditional ones rely on analytical models ~\cite{NelsonT88,lundstrom2004,tripathi2000,Ardagna:2018} and simulation~\cite{JMT,Wang2015}. Yet, recent studies have employed supervised machine learning models for performance prediction \cite{venkataraman2016ernest,MUSTAFA20183767, Hemingway, CherryPick}, which is the focus of this paper.
One such example is a regression model proposed by the Spark creators~\cite{venkataraman2016ernest}. The 
 model   uses  a reduced set of features, which are functions of the data set size and of the number of cores. The estimation of the model parameters was based on non-negative least squares.
%
%
The black-box prediction models we apply in our work exploit  variants of the features used by the {\it Ernest} model. 
Yet, instead of  non-negative least squares we use the more general $\ell_1$-regularized least squares for parameter estimation, and we also consider a set of alternative regression techniques. 

  Mustafa {\it et al.}\ \cite{MUSTAFA20183767} proposed a prediction platform for Spark SQL queries and machine learning applications, which,  similarly to our gray box models,   also exploits features related to each stage of  the Spark application. This implies the existence of previous knowledge of the  application profile. However, some of these features (e.g., numbers of \textit{nonShuffledRead}, \textit{shuffledReadRecords} and
\textit{inputPartitions}\footnote{\url{https://spark.apache.org/docs/latest/rdd-programming-guide.html}}) are at a lower level compared to ours, and thus require a finer-grained analysis of the Spark log to be computed.  The authors reported prediction errors of 10\% for SQL queries and about 25\% for machine learning jobs. 
As we will show, our approach achieves better accuracy, and our experimental design considers more recent Spark workloads,  including deep learning use cases.

CherryPick \cite{CherryPick} is  a system that leverages Bayesian optimization to find near-optimal cloud configurations that minimize cloud usage costs for MapReduce and Spark applications. Unlike other studies, the goal was not to   accurately predict applications performance, but rather design a model that is accurate enough to distinguish the best configuration from the rest. Similar ideas were also exploited in the design of Hemingway~\cite{Hemingway}, which embeds the {\it Ernest} model and is specialized in the identification of the optimal
cluster configuration for Spark MLlib based applications.  


%
%

In a related, but different direction,  Nguyen {\it et al.}  \cite{nguyen2018towards} proposed a strategy to generate training data to fit a performance model. The model is meant to be used for predicting which Spark settings yield the smallest application execution time (i.e., capacity planning). In contrast, we here compare alternative ML models and feature sets in the task of predicting the performance of an application running on a given configuration. Yet, given the significantly higher accuracy we achieve with respect to the state of art method, we argue that the proposed solution can be  beneficial for addressing the capacity planning problem as well~\cite{TCCEugenio,maros2019cloudcom}.

\section{Machine Learning Models}
\label{sec:models}

In this section, we present the proposed regression models as well as the   {\it Ernest}~\cite{venkataraman2016ernest} model, here taken as {\it reference}. Each model learns a function that estimates the execution time of an application starting from its characteristics  and from infrastructure settings. 
In this work, we consider three classes of applications:  SQL based workloads, traditional machine learning algorithms, and SparkDL (Deep Learning pipelines for Spark).
%
SparkDL relies also on TensorFlow~\cite{tensorflow2015-whitepaper} as an external library for deep network models evaluation.
Details of the applications will be presented in \autoref{sec:setup:workload}.

 We here aim at designing models that produce {\it accurate estimates of execution time}. In the future, we plan to use these estimates  to drive decisions on the best size of a cluster before it is deployed, i.e., for determining the minimum amount of resources that must be allocated to meet deadline constraints. 
Next, we present the regression techniques and the features used by each considered model. 

\subsection{Overview of Model Techniques} \label{sec:models_techniques}

The reference model, Ernest \cite{venkataraman2016ernest}, is based on linear regression, with coefficients estimated through non-negative least squares (NNLS). This is a variant of the least squares algorithm  with the added restriction that coefficients must be non-negative. This restriction ensures that each term provides a non-negative contribution to the overall execution time, preventing estimation of negative execution times.

As alternative to Ernest, we consider four classic ML models for regression: $\ell_1$-regularized linear regression (LASSO)\footnote{The $\ell_1$-regularized least squares method does not require non-negative parameters, being thus more general than the  NNLS method used in Ernest.}, neural network, decision tree, and random forest. Linear regression (LR) was chosen for being easily interpretable. Decision tree (DT) and random forest (RF), in turn, capture non-linear relationships in the data besides allowing interpretability  as well. Lastly, neural networks (NN) can capture non-trivial interactions among the input features albeit less interpretable. Investigating different models is important to analyzing the performance differences across applications and  identifying the   solution  that is best for each scenario. We refer to \cite{bishop2006pattern} for a  description of the aforementioned  regression techniques.  

\subsection{Sets of Features}
\label{sec:model:features}

Table~\ref{tab:features} shows the features used by the analyzed models. In addition to those exploited by the Ernest model, we consider two feature sets which differ by the required level of detail of the application execution. Each feature set is used as input to each of the four regression techniques presented in Section \ref{sec:models_techniques} (LR, NN, DT, RF), thus producing eight  models. We distinguish these models by the feature set used, referring to them as  black box and gray box models.  

The Ernest and the black box models use only features which are based on the number of cores and on the input data size, which are available {\it a priori} (before execution starts). The black box models use variants of the features exploited by Ernest. The data size and the number of cores are respectively the most basic information about application and infrastructure. The logarithm of the number of cores encodes the cost of reducing operations in parallel frameworks \cite{Pacheco2011}. Lastly, the ratio of data size to the number of core captures the time spent in parallel processing.

The gray box models leverage features available {\it a priori} and features only available {\it a posteriori}.
The latter are associated with the Spark DAG (directed acyclic graph), which represents the sequence of stages executed by Spark when running an application\footnote{\url{https://data-flair.training/blogs/dag-in-apache-spark/}}.
Features associated with the DAG can be extracted from the application logs after execution completion, and thus may
  be used to predict the application execution time.
In general, since the relationship between these metrics and running time only holds for the same DAG, the DAG structure of an application must be fixed across different numbers of cores to be able to build a model to predict its performance.
%
%

The information provided by the DAG was encoded as features as follows.
%
For each stage  of the DAG, we extracted the number of tasks, the maximum and the average execution time of the tasks, the maximum and the average shuffle time, the maximum and the average number of bytes transmitted among the stages. 
Such information can be easily extracted by parsing Spark logs of previous executions. 
Moreover, for SparkDL, not only the number of cores assigned to Spark executors but also
the inverse of  the total number of cores available in the cluster are included as features.
It is worth noting that our SparkDL runs were performed without the usage of GPUs to accelerate TensorFlow computation since the embedded deep network models are used for inference and are already trained. In the scenario we considered, the performance speedup achieved by GPUs would not justify the additional cost of virtual machines instances.  
%

We emphasize that while full information for all data points in the training set was used to learn the gray box models, we use only the information available before the runs to evaluate these models on the test set. Specifically, DAG-related features in the test set were replaced by their respective averages from the training data. The rationale is that when using an ML model for prediction, we cannot use these features' actual values to estimate completion time since they are only available {\it a posteriori}. 
\begin{table}[ht!]
\centering
\scriptsize
\caption{Sets of features used by different models analyzed.}
\label{tab:features}
\begin{tabular}{|c|l|}
\hline
{\textbf{Model}}                 & {\textbf{Features}} \\
\hline
\multirow{4}{*}{Ernest \cite{venkataraman2016ernest}} & - Ratio of data size to number of cores        \\
                                          & - Log of number of cores \\
                                          & - Square root of ratio of data size to number of cores \\
                                        & - Ratio of squared data size to number of cores  \\
                                        \hline
\multirow{4}{*}{Black box models} &   - Ratio of data size to number of cores \\
                                    & - Log of number of cores     \\
                                          & - Data size        \\
                                          &  - Number of cores \\
                                          & - Number of TensorFlow cores (SparkDL only) \\
\hline
\multirow{7}{*}{Gray box models} & All black box models  features  and:        \\
                           &    - Number of tasks      \\
                           &    - Max/avg time over tasks \\
                           &    - Max/avg shuffle time \\
                           &    - Max/avg number of bytes transmitted between stages\\
                           &    - Inverse of number of TensorFlow cores (SparkDL only)\\
\hline
\end{tabular}
\end{table}
\section{Experimental Setup} \label{sec:setup}

In this section, we first describe the applications used as workloads in our experiments (\autoref{sec:setup:workload}) and 
target platforms on which these applications were executed (\autoref{sec:setup:hosting}). We then discuss the splitting of data into training and test sets  (\autoref{sec:setup:split}),
model parameterization (\autoref{sec:setup:tune}) and the metrics used to evaluate  them (\autoref{sec:setup:metrics}).


%
%
%
%
%

\subsection{Workloads}
\label{sec:setup:workload}

To evaluate the accuracy of the performance prediction models, we consider three applications, which are representatives of different types of workloads: 
 a query (Query 26 from the TPC-DS industry benchmark\footnote{\url{http://www.tpc.org/tpcds}}), an ML benchmark (K-means from Sparkbench\footnote{\url{https://codait.github.io/spark-bench}}), and an ad-hoc benchmark for image processing we developed based on SparkDL library \footnote{\url{https://github.com/databricks/spark-deep-learning.}}. We ran each application  in various scenarios, as shown below, to build  our training and test sets. 
 
%
%
Query 26 is an interactive query, representative of SQL-like workloads, which includes a small number of tasks and stages (i.e., 10).
It was run for various input data set sizes: 250 GB, 750 GB, and 1000 GB. 

The K-means clustering algorithm   is the core of many ML applications. It is an unsupervised learning technique that is used on its own but also to perform preliminary analyses in some classification tasks.
As other ML algorithms,  K-means is an iterative workload, usually characterized by larger execution time variability.
It was run for Spark dataframes with 100 features, with values uniformly distributed in  the [0,1] range and the number of rows varying in 5, 10, 15 and 20 million. We found that the DAG was the same for all the data sizes, containing exactly 15 stages.

Lastly, the SparkDL based workload  is an example of a high-level deep learning application on Spark. It is a binary image classification using \textit{InceptionV3} as featurizer and \textit{linear SVM} as classifier. 
The number of images in the input varied in 1000, 1500 and 2500 while the number of stages is~8.
As anticipated in \autoref{sec:model:features}, the SparkDL benchmark is characterized by additional features and, hence, is the most complex of the three considered workloads.
SparkDL heavily relies on TensorFlow which affects the application completion times.
When SparkDL runs, the number of cores allocated to Spark workers can be limited, but \textit{spark-submit} parameters cannot control the number of cores for TensorFlow, which uses all the cores available in the cluster.
For this reason, also the TensorFlow number of cores (which corresponds to the number of cores available in the cluster) was included in the  feature set of both black and gray box models while  its inverse is used in the gray box models only.

\subsection{Hosting Platforms}
\label{sec:setup:hosting}

We ran Spark applications on two platforms, Microsoft Azure and a private IBM Power8 cluster, which are representatives of different computing environments.
As a public cloud, Microsoft Azure is potentially affected by resource contention. Thus, application executions might experience more variability.
In contrast, IBM Power8 was fully dedicated to run our benchmarks without any other concurrent activity (thus with no resource contention).
%

Query 26 and SparkDL were executed on Microsoft Azure using the HDInsight service with workers based on 6 D13v2 virtual machines (VMs), each with 8 CPU cores and 56 GB of memory running Spark 2.2.0 on Linux.
SparkDL application requires, in addition, that TensorFlow and Keras are available on the Spark cluster: versions 1.4.0 and 2.1.5 were used, respectively.
The   executors memory was set  to 10 GB.
K-means was run on a Power8 deployment that includes Hortonworks distribution 2.6, same as Microsoft Azure, with 4 VMs, each with 12 cores and 58 GB of RAM, for a total of 48 CPU cores available for Spark workers, plus a master node with 4 cores and 48 GB of RAM. 
The executors memory, in this case, was set  to 4GB.

For Query 26 and K-means, we ran experiments varying the number of Spark cores between 6 and 44 cores (step of 2), repeating the execution with the same configuration  6 times.
%
%
For SparkDL, we varied the number of cores between 2 and 48 (step of 2), repeating each experiment with the same configuration (i.e., the number of images and cores) 5 times.
By considering different workloads, hosting platforms and setup configurations, we build a rich set of scenarios to test our prediction models.

\subsection{Training and Test Sets}
\label{sec:setup:split}

\begin{figure*}[!tttt]
\centering
\begin{subfigure}[b]{0.26\linewidth}
    	\includegraphics[width=\linewidth]{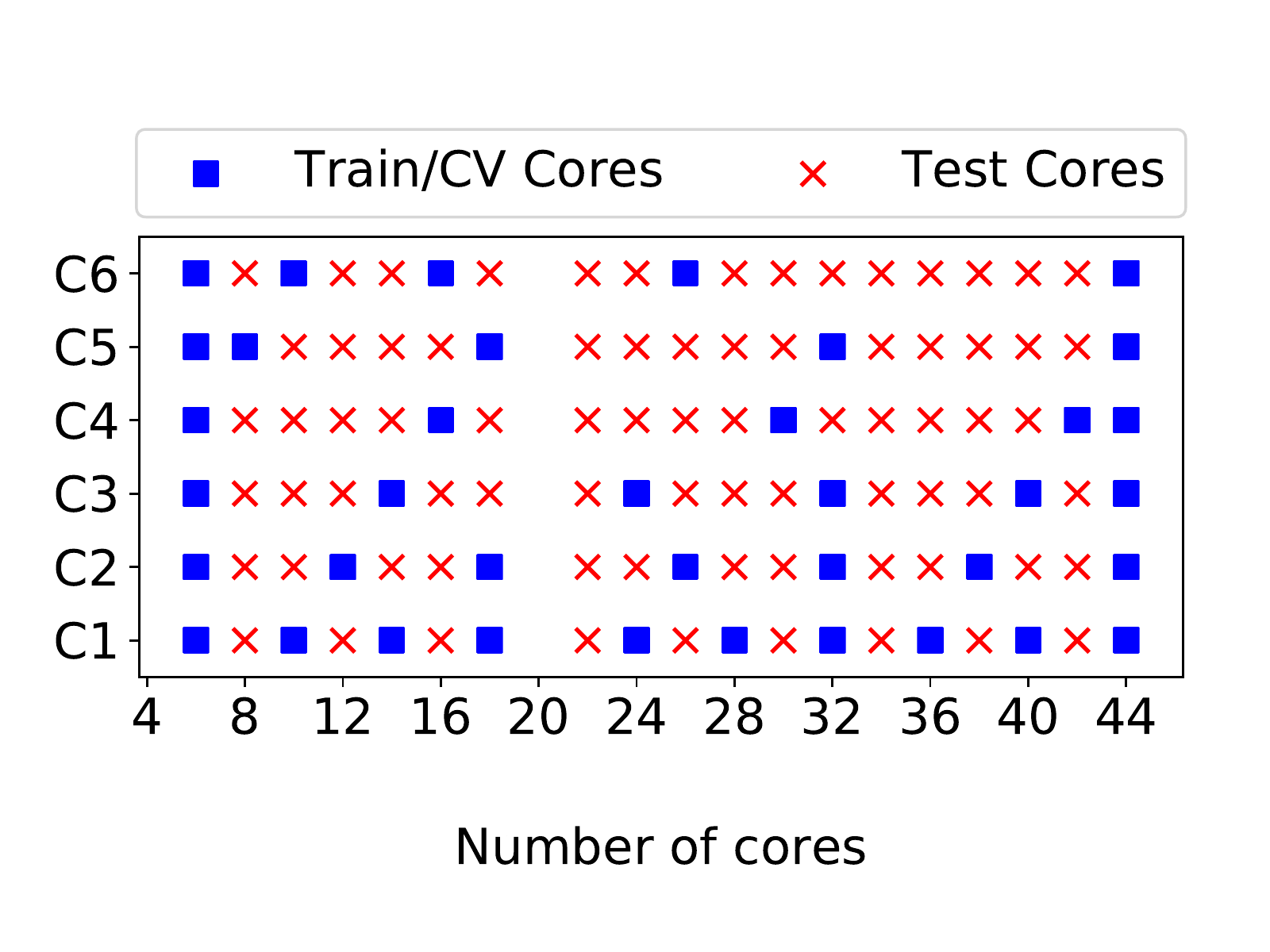}
    	\caption{Query 26}
	\label{fig:cases_for_queries}%
\end{subfigure}
\begin{subfigure}[b]{0.26\linewidth}
    	\includegraphics[width=\linewidth]{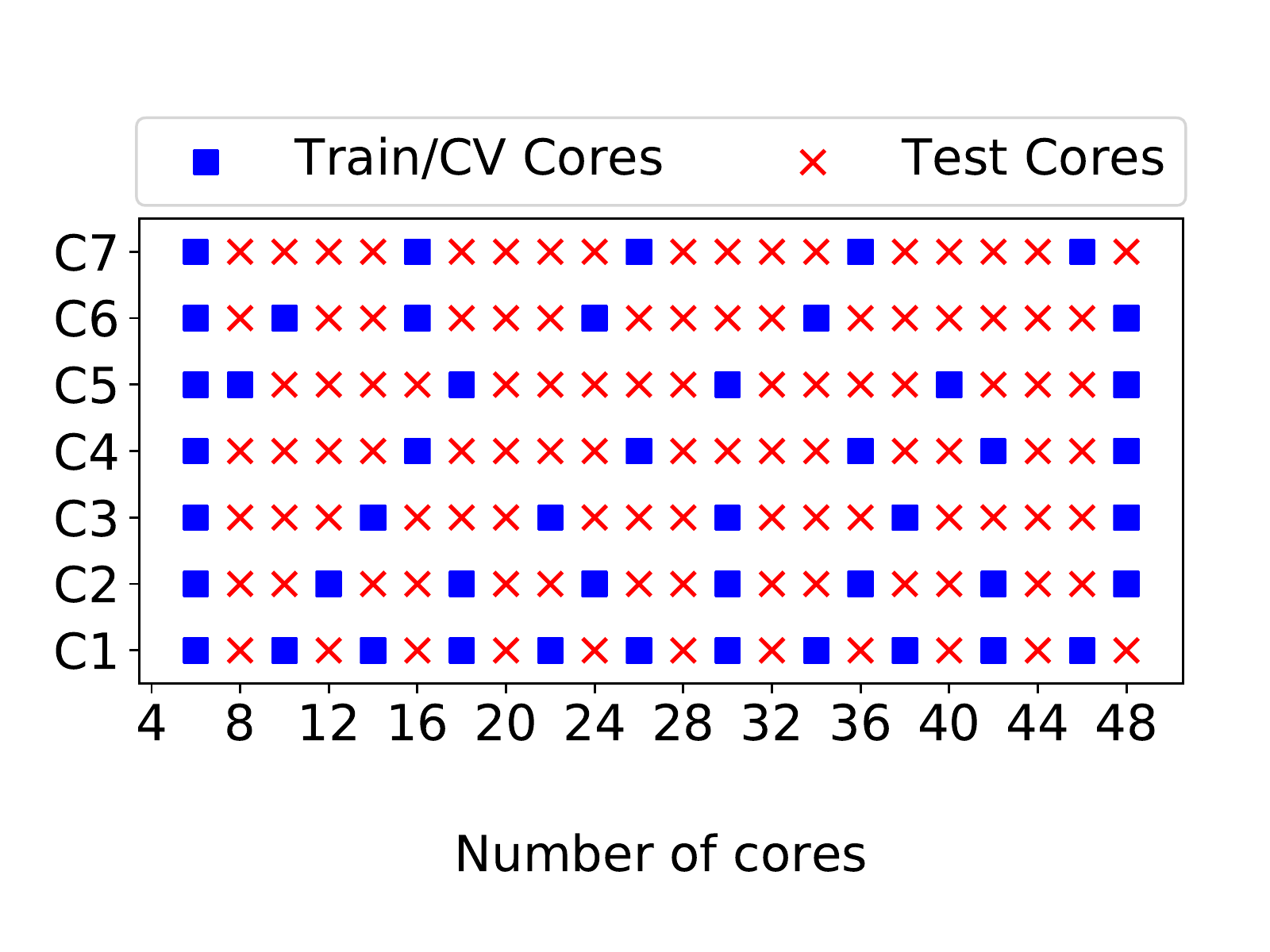}
   	\caption{K-means}
	\label{fig:cases_for_others}%
\end{subfigure}
\begin{subfigure}[b]{0.26\linewidth}
    	\includegraphics[width=\linewidth]{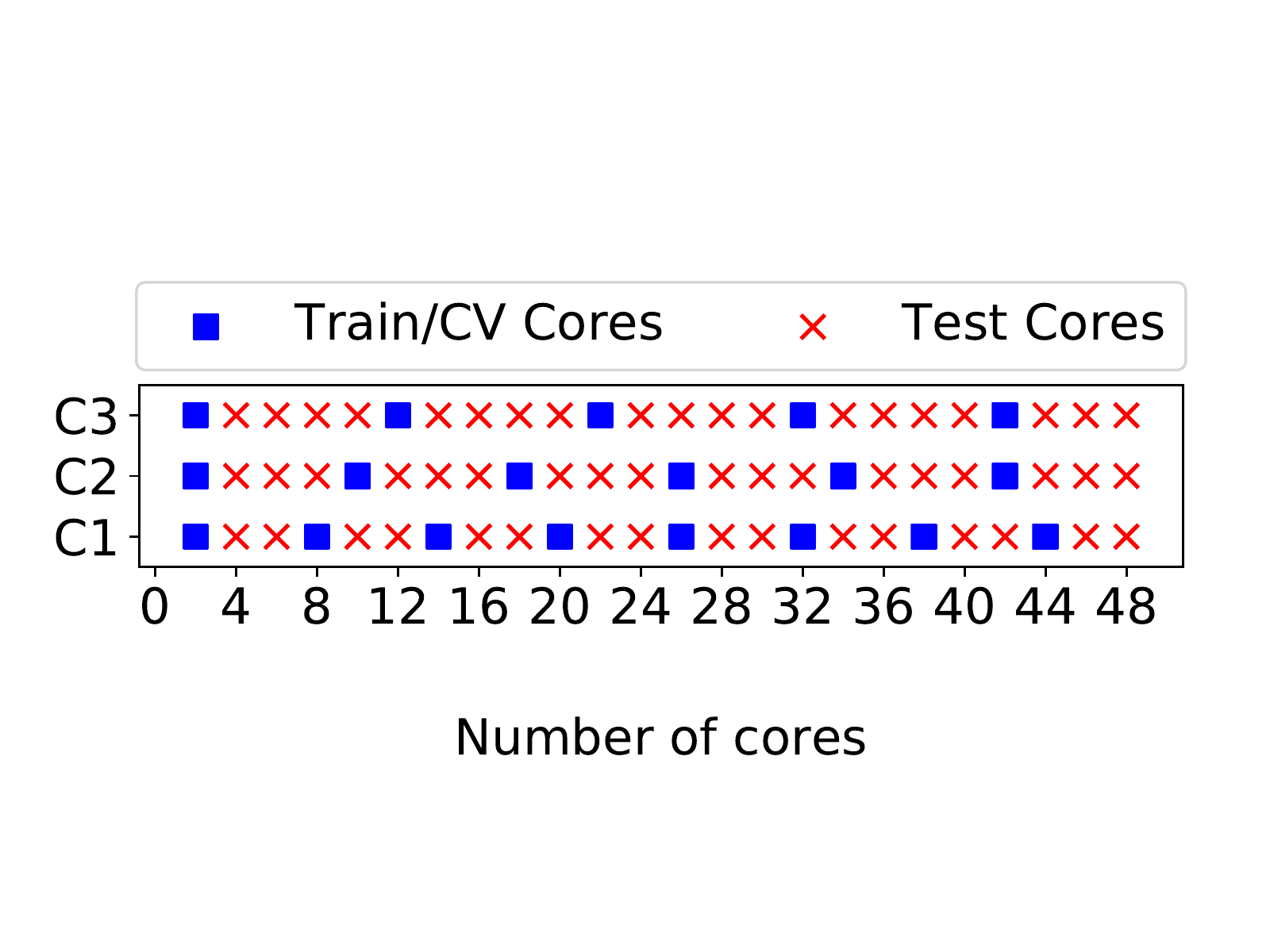}
    	\caption{SparkDL}
	\label{fig:cases_for_runbest}%
\end{subfigure}
\caption{Core interpolation scenario: train-test split in each case for Query 26, K-means and SparkDL.} 
\label{fig:cases_for_all}
\end{figure*}

To learn and evaluate the ML models, data coming from the experiments  are split into training and test sets. The former is used to learn  the model, whereas the latter is used to evaluate its accuracy.  Since hyper-parameters tuning is performed for each ML technique (see \autoref{sec:setup:tune}), a subset of the training  data is used for cross-validation to reduce over-fitting.
For each workload, we evaluate the accuracy of the prediction models in terms of {\it core interpolation} and {\it data size extrapolation} capabilities, acknowledging the fact that the data available for learning the models (training set) might have been obtained via experiments on setups different from the one for which the prediction is needed (test set).
Figure \ref{fig:cases_for_all} and Table \ref{tab:data_size} summarize the scenarios considered. 
%

%

In the \emph{core interpolation} scenarios, we consider runs with the same dataset size (reported in \autoref{tab:data_size}) and   verify the capabilities of the trained models of interpolating the number of cores.
Figure \ref{fig:cases_for_all} shows the various scenarios (y-axis) of {\it core interpolation} built for each workload based on different splits of the data into training and test sets: in each row (case number), blue boxes represent configurations for which the data were used as part of the training set (and cross-validation) and red crosses indicate configurations used as part of the test set\footnote{Since the experiments with Query 26 on 20 cores presented some anomalies, we chose to remove them from the training and testing data (see Figure \ref{fig:cases_for_queries}).}. 
We designed scenarios  such that larger case numbers are associated with harder predictions, as their training data include samples from a smaller range of experiments w.r.t. the number of cores. 
%

%
%
%
%
%

%
%
For example, for both Query 26 and K-means, scenario C1  is built by alternating configurations in the sequence of the number of cores (x-axis) as training and test data. For Query 26, 
data from experiments with the number of cores equal to 6, 10, \ldots, 40 and 44 are put in the training data (blue boxes) while the remaining samples are included in the test set (red cross).
We gradually incremented the gap between the number of cores of consecutive configurations included in the training data. We varied the gap in cases 4, 5, 6, and 7 to assess its impact on model accuracy. Since there is a large difference in the application completion time in the runs where the number of cores is small, we always included the data for the smallest number of cores in the training set.  
 For SparkDL, we proceeded along the same lines but limiting the number of cases to three.
%
%
%
%

%

In the \emph{data size extrapolation} scenarios, we put the runs with the largest dataset size (spanning across all available cores configurations) in the test set while the runs with the other dataset sizes in the training data, as shown in the two rightmost columns of \autoref{tab:data_size}.
%
%
Moreover, training sets are further reduced by removing runs 
according to the same schema presented for core interpolation.
By doing so, in these experiments we evaluate  at the same time the core interpolation and the data  size extrapolation capabilities.  In other words,
these experiments differ from the core interpolation scenarios because: (i) the dataset sizes in training and test sets are no longer the same, (ii) in the test set we also include observations where the number of cores is the same as in some observations of the training set (but again with different dataset sizes).
%

\begin{table} [ttt]
    \centering
    \vspace{-0.2cm}
    \caption{Workload data sizes in different scenarios}
    \scriptsize
    \begin{tabular}{|c|c|c|c|c|}
    \hline
             \multirow{2}{*}{Workload} & \multicolumn{2}{c|}{Core Interpolation} & \multicolumn{2}{c|}{Data Extrapolation}\\
             \cline{2-5}
     & Training & Test & Training & Test \\
             \hline
         Query 26 [GB] & 750 & 750 & 250, 750 & 1000 \\
         K-means [Rows] & 15 & 15 & 5, 10, 15 & 20 \\
         SparkDL [Images] & 1500 & 1500 & 1000, 1500 & 2000 \\
             \hline
    \end{tabular}
        \vspace{-0.2cm}
    \label{tab:data_size}
\end{table}

\subsection{Hyper-parameter Tuning}
\label{sec:setup:tune}
%
An inhouse Python library was developed on the top of PyTorch 0.4.0\footnote{https://pytorch.org} (for neural network training) and of scikit-learn 0.19.1\footnote{https://scikit-learn.org} (for all the other techniques) to explore different values of hyper-parameters as shown in Table \ref{tab:hyperparam}. For every algorithm, we used the  Mean Squared Error (MSE) to select the  values that led to the best model. The hyper-parameters more frequently used are reported in bold in the table.  For black box models, this evaluation was done via \textit{5-fold} cross-validation, while gray box models, which  require much longer training times due to the larger feature set,  were parameterized based on \textit{hold-out} cross-validation.
%

To prevent over-fitting, a regularization term is added to LR (Linear Regression) and NN (Neural Network).
For LR, LASSO was chosen providing $\ell_1$-norm.  The use of intercept and the penalty constant $\alpha$  are the set of available hyper-parameters and are shown in~\autoref{tab:hyperparam}

\begin{table}[h!]
\centering
\caption{Hyper-parameters for ML techniques} \label{tab:hyperparam} 
\scriptsize
\begin{tabular}{|l|l|l|}
\hline
\multicolumn{3}{|c|}{{\bf Linear Regression (LR)}} \\  \hline
\textbf{Hyper-Parameter} & \multicolumn{2}{|c|}{\textbf{Values}} \\ \hline
Penalty $\alpha$       & \multicolumn{2}{|c|}{0.001, 0.01, 0.05, 0.1, 0.5, {\bf 1.0}, 5.0, 10.0}    \\
Fit intercept & \multicolumn{2}{|c|}{{\bf True}, False} \\ 
\hline \hline
\multicolumn{3}{|c|}{{\bf Neural Network (NN)}} \\  \hline
\textbf{Hyper-Parameter} & \multicolumn{2}{|c|}{\textbf{Values}} \\ \hline
\# Layers $n$ & \multicolumn{2}{|c|}{1, {\bf 2}, 3} \\
\# Perceptrons/Layer & \multicolumn{2}{|c|}{all combinations in $[3, 4, \mathbf{5}]^n$ }\\
Activation Functions & \multicolumn{2}{|c|}{sigmoid, {\bf ReLU}, tanh} \\
$\ell_2$ Penalty &\multicolumn{2}{|c|}{ 0.0001, {\bf 0.001}, 0.01, 0.05, 0.1} \\
Learning Rate & \multicolumn{2}{|c|}{0.001, {\bf 0.01}, 0.1} \\
$\beta_1$ & \multicolumn{2}{|c|}{0.7, 0.8, {\bf 0.9}} \\
\# Minibatches & \multicolumn{2}{|c|}{{\bf 1}} \\
Optimizer & \multicolumn{2}{|c|}{{\bf Adam}, SGD}\\
\hline \hline
\multicolumn{3}{|c|}{{\bf Decision Tree (DT) \& Random Forest (RF)}} \\  \hline
\textbf{Hyper-Parameter} & \textbf{Decision Tree} & \textbf{Random Forest} \\ \hline
	Max Depth & 3, {\bf 5}, 10, No Limit & 3, 10, {\bf 20}, No Limit \\
	Max Features & {\bf auto}, sqrt, log & auto, {\bf sqrt}, log \\
	Min samples to split & {\bf 2.0} & {\bf 2.0} \\
	Min samples per leaf & {\bf 1\%}, 5\%, 10\%, 20\%, 30\% & 1, {\bf 2}, 4 \\
	Criterion & {\bf MSE}, fMSE, MAE & {\bf MSE}, MAE \\
	\# Trees & NA & 5, {\bf 10}, 50, 100 \\
\hline
\end{tabular}
\end{table}

For the NN algorithm, the $\ell_2$ penalty Frobenius norm \cite{GoluVanl96} was used. Also   rectified linear unit function \emph{ReLU} was selected in general as the best activation function. 
In all cases, the training data size was not large; therefore, we set the number of minibatches to 1 in order to consume the whole input at once.
The main hyper-parameter considered to evaluate the performance was the optimizer.
\emph{Adam} and \emph{Stochastic Gradient Descent (SGD)} were evaluated as possible candidates. 
%
The former was selected as our experiments showed  that it converges much faster than the latter. The number of epochs was set to 10,000.

DT and RF share many hyper-parameters and their values, therefore, are grouped in~\autoref{tab:hyperparam}. \emph{Max Depth} is the maximum depth of the tree which is specified to avoid over-fitting. \emph{Max Features} is used to select the number of available features to consider when searching for the best split. A value of \emph{auto} implies a maximum number of features equal to the total number of features.
Values of    \emph{sqrt} and  \emph{log} imply a maximum  equal to   the square root and the base-2 logarithm of the number of features, respectively. \emph{Minimum Samples to Split/per Leaf} is used for setting, respectively, the minimum number of samples required to split a node and the minimum percentage/number of samples required to be a leaf. \emph{Criterion} is the function used to measure the quality of a split:  \emph{MSE} stands for mean square error (minimizes $\ell_2$ loss), \emph{fMSE} stands for mean squared error with Friedman's improvement score for potential splits and, last, \emph{MAE} stands for mean absolute error (minimizes $\ell_1$ loss). 
Parameter \emph{number of trees},    the number of trees in the forest, only applies to RF. A range of values is explored to analyze diminishing return effects on the error.

\subsection{Performance Metrics}
\label{sec:setup:metrics}

We  evaluate the performance of each model based on the mean absolute percent error (MAPE) of the predicted response time, as it is more  widely used in the performance literature \cite{Lazowska} than MSE (used for hyper-parameter tuning). Also, it allows us to consolidate results across experiments with different execution times. MAPE measures the relative error (in absolute terms) of the prediction with respect to the true response times, i.e.,  $\textrm{MAPE (\%)} = {\frac{100}{N} \sum\limits_{k=1}^N {\left|\frac{y_k\ -\ \hat y_k}{y_k}\right|}}$, where $N$ is the number of data points, $y_k$ is the  response time measured on the operating system, and  $\hat y_k$ is the predicted response time from the learned model. For each setup, 10 runs were executed for the LR, DT and RF algorithms, and  average MAPE across all 10 runs are reported. For NN, which has much longer training times, we performed a single run (i.e., random train-test split). For each of the gray box models (which include all features from black box models plus some others), Table \ref{tab:time} reports the average time to perform the full training campaign (including the hyper-parameter tuning phase). Experiments were run on a Ubuntu 18.04 Virtual Machine with 8 cores and 10 GB of memory hosted within an Intel Xeon Silver 4114 Ubuntu 16.04  server with 20 cores and 48 GB of memory.

\begin{table}[t!]
\centering
\caption{ML Models training times (in minutes)} \label{tab:time} 
\scriptsize
\begin{tabular}{|l|c|c|c|c|}
\cline{2-5}
 \multicolumn{1}{c|}{} & LR & NN & DT & RF  \\
\hline
training time & 1 & 908 & 3 & 56 \\
\hline
\end{tabular}
\end{table}

%
%

%

%


\section{Experimental Results}\label{sec:results}

In this section, we present our prediction results for each workload  on  two sets of scenarios, namely  \emph{core interpolation} and \emph{data extrapolation}. For each set, we report results for  every case described in \autoref{sec:setup:split}. We discuss the accuracy of the models described in \autoref{sec:models}, referring to each of them as gray box or black box, and, within each category,  specifying the ML technique employed (LR, NN, DT or RF).  We compare them against the  reference model (Ernest) with respect to MAPE  on the test set. 
\subsection{Results for Query 26}

Tables~\ref{tab:mape_q26_power_8_simplified_configuration_2} and \ref{tab:mape_q26_power_8_simplified_configuration_4} show the MAPE results for core interpolation and data extrapolation scenarios, respectively, for Query 26. In both cases, we generally observe that: (i) MAPE of black box models is smaller than that of their gray box counterparts, (ii) Ernest performs relatively well -- MAPE within $1.5-1.7\%$ for interpolation and within $7.3-8.0\%$ for extrapolation, and (iii) best results are obtained by black box LR model, which achieves even smaller MAPE values than Ernest. DT and RF yield larger MAPE on the extrapolation scenarios, which can be explained by the fact that their execution time predictions are based on the averages of observed values from the training set.

\begin{table}[!ttt]
\centering
\scriptsize
\captionof{table}{MAPE (\%) of execution time estimates on Power8 for Query 26 (core interpolation; fixed data size of 750 GB for all datasets).}   \label{tab:mape_q26_power_8_simplified_configuration_2}
\begin{tabular}{|l||c|c|c|c||c|c|c|c||c|}
\hline
\multirow{2}{*}{} & \multicolumn{4}{c||}{\bfseries Gray Box Models} & \multicolumn{4}{c||}{\bfseries Black Box Models} & \multirow{2}{*}{\bfseries Ernest} \\
\cline{2-9}
& DT & LR & NN & RF & DT & LR & NN & RF & \\
\hline
C1  & 20.6 &  63.4 & 12.3 & 18.8 & 8.4  & {\bf 1.0} & 6.7  & 2.4  & 1.5 \\
C2  & 16.7 &  72.6 & 16.1 & 19.9 & 7.9  & {\bf 1.2} & 20.1 & 6.0  & 1.6 \\
C3  & 18.0 &  98.5 & 36.3 & 18.9 & 11.2 & {\bf 1.2} & 3.1 & 8.0  & 1.7 \\
C4  & 21.7 & 300.6 & 18.9 & 27.2 & 12.9 & {\bf 1.1} & 4.4  & 9.7  & 1.6 \\
C5  & 35.7 & 229.7 & 30.8 & 35.0 & 12.5 & {\bf 1.2} & 23.0 & 12.0 & 1.6 \\
C6  & 27.0 & 414.1 & 26.3 & 32.3 & 8.9  & {\bf 1.2} & 5.4  & 6.9  & 1.7 \\
\hline
\end{tabular}
\end{table}

\begin{table}[!ttt]
\centering
\scriptsize
\captionof{table}{MAPE (\%) of execution time estimates on Power8 for Query 26 (data extrapolation; 250 GB, 750GB for training and 1000GB for testing).}   
\begin{tabular}{|l||c|c|c|c||c|c|c|c||c|}
\hline
\multirow{2}{*}{} & \multicolumn{4}{c||}{\bfseries Gray Box Models} & \multicolumn{4}{c||}{\bfseries Black Box Models} & \multirow{2}{*}{\bfseries Ernest} \\
\cline{2-9}
& DT & LR & NN & RF & DT & LR & NN & RF & \\
\hline
C1 & 38.2 & 28.6 &  7.0 & 39.6 & 15.6 & {\bf 3.9} & 9.9  & 19.4 & 7.5 \\
C2 & 42.5 & 23.5 & 24.8 & 33.6 & 16.0 & {\bf 4.0} & 10.3 & 16.6 & 7.4 \\
C3 & 39.0 & 36.7 & 11.2 & 37.2 & 21.1 & {\bf 3.7} & 7.8  & 19.0 & 7.3 \\
C4 & 42.2 & 33.3 & 13.5 & 35.4 & 25.5 & {\bf 4.0} & 25.4 & 23.5 & 7.3 \\
C5 & 32.5 & 12.4 &  9.8 & 35.1 & 19.0 & {\bf 4.4} & 31.8 & 17.3 & 7.6 \\
C6 & 37.0 & 24.8 & 24.8 & 37.3 & 17.3 & {\bf 4.9} & 18.1 & 19.5 & 8.0 \\
\hline
\end{tabular}
\label{tab:mape_q26_power_8_simplified_configuration_4}
\end{table}

\subsection{Results for K-means }

Tables~\ref{tab:mape_k-means_simplified_configuration_3} and \ref{tab:mape_k-means_simplified_configuration_5} show the results for core interpolation and data extrapolation scenarios, respectively, for K-means. In contrast to the results  for Query 26, we generally observe that (i) Ernest performs extremely poorly in both sets of scenarios, (ii) for interpolation, black box RF achieves the best results, and (iii) for extrapolation, gray box LR outperforms its black box counterpart more often than not. Although DT and RF achieve the best results on extrapolation, this may not hold true in cases when the data size of observations in the test set is much larger (or smaller) than that of the observations in the training set (i.e., differences are more extreme than those than captured by our scenarios).
In those cases, we expect the other models to outperform DT and RF.

\begin{table}[!ttt]
\centering
\scriptsize
\captionof{table}{MAPE (\%) of execution time estimates on Power8 for K-means  (core interpolation; 15M points for training and 15M points for test).}   \label{tab:mape_k-means_simplified_configuration_3}
\begin{tabular}{|l||c|c|c|c||c|c|c|c||c|}
\hline
\multirow{2}{*}{} & \multicolumn{4}{c||}{\bfseries Gray Box Models} & \multicolumn{4}{c||}{\bfseries Black Box Models} & \multirow{2}{*}{\bfseries Ernest} \\
\cline{2-9}
& DT & LR & NN & RF & DT & LR & NN & RF & \\
\hline
C1 & 27.7 & 184.3 &   77.9 & 24.8 & 16.0 & 50.8 & 38.3 & {\bf 5.0}  & 126.7 \\
C2 & 28.7 & 225.0 &  109.6 & 54.7 & 16.9 & 46.3 & 18.1 & {\bf 13.7} & 148.1 \\
C3 & 22.9 & 278.3 &  435.7 & 26.0 & 18.5 & 42.1 & 10.3 & {\bf 14.0} & 161.3 \\
C4 & 33.8 & 300.6 &  445.1 & 26.3 & 21.4 & 41.9 & 23.6 & {\bf 14.2} & 176.5 \\
C5 & 27.1 & 543.4 & 1146.1 & 22.5 & 22.9 & 42.3 & 31.3 & {\bf 19.3} & 187.0 \\
C6 & 33.9 & 414.1 &  170.8 & 91.3 & 15.2 & 48.2 & {\bf 10.6} & 12.0 & 159.9 \\
C7 & 22.6 & 363.1 &  626.0 & 31.3 & 17.4 & 41.8 & 33.2 & {\bf 14.8} & 178.1 \\
\hline
\end{tabular}
\end{table}

\begin{table}[!ttt]
\centering
\scriptsize
\captionof{table}{MAPE (\%) of execution time estimates on Power8 for K-means  (data extrapolation; 5M, 10M, and 15M points for training and 20M points for test).}   
\begin{tabular}{|l||c|c|c|c||c|c|c|c||c|}
\hline
\multirow{2}{*}{} & \multicolumn{4}{c||}{\bfseries Gray Box Models} & \multicolumn{4}{c||}{\bfseries Black Box Models} & \multirow{2}{*}{\bfseries Ernest} \\
\cline{2-9}
& DT & LR & NN & RF & DT & LR & NN & RF & \\
\hline
C1 & 24.0 & 20.2 &  22.9 & 27.1 & 19.8 & 40.0 & 32.5 & {\bf 12.5} &  93.4 \\
C2 & 35.6 & 16.0 & 121.3 & 20.7 & {\bf 14.5} & 39.6 & 29.1 & 16.5 & 107.8 \\
C3 & 66.7 & 31.9 & 134.1 & 32.6 & {\bf 14.4} & 41.4 & 28.0 & 16.5 & 119.6 \\
C4 & 28.8 & 24.2 & 118.9 & 25.8 & {\bf 13.3} & 31.8 & 68.4 & 16.0 & 127.9 \\
C5 & 42.0 & 34.0 &  27.5 & 26.7 & {\bf 15.3} & 25.7 & 60.0 & 19.5 & 121.4 \\
C6 & 75.7 & 37.7 &  37.1 & 24.2 & {\bf 11.0} & 52.6 & 26.7 & 14.9 & 109.8 \\
C7 & 32.6 & 43.1 & 148.1 & 42.6 & 20.8 & 34.5 & 96.4 & \bf{17.9} & 129.5 \\
\hline
\end{tabular}
\label{tab:mape_k-means_simplified_configuration_5}
\vspace{-0.2cm}
\end{table}

\subsection{Results for SparkDL}

Finally, results for SparkDL are shown in Tables~\ref{tab:mape_runbest_simplified_configuration_2} and \ref{tab:mape_runbest_simplified_configuration_4} for the core interpolation and data extrapolation scenarios, respectively. For interpolation, the black box DT and RF tend to yield the best predictions, although LR also performs well, achieving smaller MAPE than Ernest. The gray box models do not provide significant improvements (except for a few cases) but often incurs great degradation in MAPE. In contrast, for extrapolation, the additional features estimated by gray box models combined with non-linear models (notably NN and RF) very often improve upon their black box counterparts. However, the black box LR is still the model with best performance, achieving a  MAPE  roughly 5 times smaller than  that obtained by Ernest.


\section{Discussion} \label{sec:discussion}


The previous section presented the prediction accuracy of each model on three quite different workloads for various scenarios. 
Ernest performs well (MAPE $<$ 11\%) in the simplest scenarios considered (i.e., Query 26 and core interpolation for SparkDL). Yet, in the other scenarios, the large MAPE values yielded by the model make it unsuitable for production environments.
In fact, in the worst case, the error reaches up to 187\%, showcasing the need for new techniques to overcome Ernest's limitations.
%
%
Our goal was to compare alternative techniques against the simple linear regression (with NNLS estimation) applied by Ernest, investigating the extent to which the additional information (i.e., features) used by gray box models   increases prediction accuracy. 
%
%

\begin{table}[!tt]
\centering
\scriptsize
\captionof{table}{MAPE (\%) of execution time estimates on Azure for SparkDL    (core interpolation; 1500 images for training and 1500 images for testing).}   
\label{tab:mape_runbest_simplified_configuration_2}
\begin{tabular}{|l||c|c|c|c||c|c|c|c||c|}
\hline
\multirow{2}{*}{} & \multicolumn{4}{c||}{\bfseries Gray Box Models} & \multicolumn{4}{c||}{\bfseries Black Box Models} & \multirow{2}{*}{\bfseries Ernest} \\
\cline{2-9}
& DT & LR & NN & RF & DT & LR & NN & RF & \\
\hline
C1 & 5.2 & 28.7 &  4.7 &  4.5 & 5.1 & 7.3 & {\bf 4.6} & 5.1 & 10.5 \\
C2 & 5.8 &  5.7 & 13.3 &  4.8 & {\bf 5.5} & 6.2 & 8.6 & 5.7 &  6.3 \\
C3 & 8.9 &  7.5 &  5.4 &  6.0 & 5.5 & 5.5 & 5.7 & {\bf 4.9} &  5.7 \\
		\hline
\end{tabular}
\end{table}

\begin{table}[!ttt]
\centering
\scriptsize
\captionof{table}{MAPE (\%) of execution time estimates on Azure for SparkDL    (data extrapolation; 1000 and 1500 images for training and 2500 images for testing).}   \label{tab:mape_runbest_simplified_configuration_4}
\begin{tabular}{|l||c|c|c|c||c|c|c|c||c|}
\hline
\multirow{2}{*}{} & \multicolumn{4}{c||}{\bfseries Gray Box Models} & \multicolumn{4}{c||}{\bfseries Black Box Models} & \multirow{2}{*}{\bfseries Ernest} \\
\cline{2-9}
& DT & LR & NN & RF & DT & LR & NN & RF & \\
\hline
C1 & 37.0 & 10.7 & 25.7 & 34.7 & 35.9 & {\bf 7.5} & 34.1 & 36.7 & 43.5 \\
C2 & 36.3 & 10.0 & 31.4 & 37.0 & 41.5 & {\bf 7.6} & 15.3 & 41.9 & 37.4 \\
C3 & 36.9 & 14.7 &  9.9 & 34.5 & 41.0 & {\bf 7.8} & 33.3 & 41.1 & 36.8 \\
		\hline
\end{tabular}
\vspace{-0.2cm}
\end{table}

From the previous results, we observe that no single ML technique outperforms all the others in every scenario. 
Moreover, even a slight  change in the composition of training and test sets (i.e., considering a different case number) may impact the  technique that performs best.
For example, in the data extrapolation experiments with K-means, the best gray box model is the one with  LR as regression technique in cases C1, C2, C3 and C4, with RF in  cases C5 and C6, and with DT in the last case (C7). Similarly, the best black box model is obtained with RF in   cases  C1 and C7,  and   DT in the remaining five cases.  

The comparison between the best gray box models and the reference Ernest model leads to two different situations.
In scenarios where applications are characterized by regularity (i.e., Query 26 and SparkDL with fixed data size), Ernest yields very good results with MAPE values smaller than 10\%, whereas the best gray box model generally achieves worse performance (MAPE of best models is in the range $7.0$-$30.8\%$).
Yet, in the remaining scenarios, which are characterized by a larger variability in the application execution times, the best gray box model outperforms the Ernest model by a large margin.
The MAPE range of the latter is $37.4$-$187.0\%$ while the largest error of the best gray box model is only $33.9\%$ (C6 of  core interpolation with K-means). 

However, recall that gray box models do use   DAG-related  features  which are not available for the test instance at prediction time ({\it a priori}), and thus are replaced by  the averages  from  the  training  data.
Thus, even though the gray box models are able to outperform Ernest in some scenarios, such feature approximations may indeed cause accuracy degradation, as it hides significant differences that may exist between runs in the training and test sets. 
%
%
%
 Black box models, in turn,  extract features only from information available before an application starts execution (i.e., available at prediction time), and thus do not suffer from this issue.
Also, our results show that the best black model always outperforms the results of the Ernest method (possibly due to the use of more effective ML techniques)
and almost always outperforms the results of the gray box model, despite using fewer features. Thus, this approach is preferable over the others.
In the worst case (C5 of  core interpolation with K-Means) the MAPE of the best black box model is only 19.3\%, which is quite suitable  in a production environment.

Finally, the  black box LR model produces the best results for Query 26, in spite of Ernest's good performance on this application. Indeed, the black box LR model outperforms Ernest in all scenarios analyzed, which indicates the benefits of using a different parameterization technique as well as slightly different  feature set.
However, there are    scenarios where Ernest performs well (i.e.,  core interpolation with SparkDL), for which the best black box models are indeed DT and RF, though the black box LR performs quite similarly. 
%
The latter is indeed the best model also in  data extrapolation scenario with SparkDL: it achieves MAPE values in the range $7.5$-$7.8\%$, while Ernest yields values in the range $36.8$-$43.5\%$.
Yet, there is no clear winner in the K-means experiments. For core interpolation, the best results are produced with RF in all cases but C6, for which NN is the best performer. For data extrapolation,  the best technique is DT in 5 out of 7 cases and RF in the others.

\section{Conclusions and Future Work} \label{sec:conclusions}

In this paper, we investigated the
accuracy of alternative supervised ML techniques and different feature sets in the performance prediction of Spark applications. 
Experimental results on a rich set of different scenarios  demonstrated that our black box models are able to achieve at least the same accuracy of Ernest and, in many  scenarios, even more accurate predictions.
The percentage error is reduced from 126.7-187.0\% to only 5.0-19.3\% when applications present irregular patterns and/or when it is needed to extrapolate the application behavior on larger data sets. 
However, we also  showed  that there is not a single ML technique that always outperforms the others, hence different techniques have to be evaluated in each scenario to choose the best model.
As future work, we plan to study the performance of Spark deep learning applications when GPU-based clusters are used and develop capacity planning solutions to identify, at deployment time, the minimum cost cluster configuration  so as to guarantee application runs within an a priori deadline.







\section*{Acknowledgements}
{\footnotesize
This work has been partially supported by the project ATMOSPHERE (\url{https://atmosphere-eubrazil.eu}), funded by the Brazilian Ministry of Science, Technology and Innovation (Project 51119 - MCTI/RNP 4th Coordinated Call) and by the European Commission under the Cooperation Programme, Horizon 2020 (grant agreement no 777154).

\bibliographystyle{IEEEtran}
\bibliography{reference,IEEEabrv}
}

%

\end{document}